\documentclass{article}
\usepackage{graphicx} 
\usepackage{amsmath, amssymb}
\usepackage{booktabs}
\usepackage{tikz}
\usepackage{xspace}
\usepackage[utf8]{inputenc}
\usepackage{cite} 
\usepackage{makecell}
\usepackage{xcolor}
\usepackage{siunitx}
\usepackage{hyperref}
\newcommand{\ReCom}{{\sf ReCom}\xspace } 
\newcommand{\RevReCom}{{\sf RevReCom}\xspace }
\newcommand{\ReComA}{{\sf ReCom-A}\xspace } 
\newcommand{\ReComB}{{\sf ReCom-B}\xspace }
\newcommand{\ReComC}{{\sf ReCom-C}\xspace }
\newcommand{\ReComD}{{\sf ReCom-D}\xspace }
\newcommand{\RA}{{\sf A}\xspace}
\newcommand{\RB}{{\sf B}\xspace}
\newcommand{\RC}{{\sf C}\xspace}
\newcommand{\RD}{{\sf D}\xspace}
\newcommand{\popp}{\ensuremath{\text{\sf Pop}_+}}
\newcommand{\popm}{\ensuremath{\text{\sf Pop}_-}}
\newcommand{\dem}{d_{\text{em}}}
\newcommand{\dks}{d_{\text{KS}}}
\newcommand{\dosp}{d_{\text{OSP}}}

\newcommand{\ESS}{\ensuremath{n_{\text{eff}}}}

\newcommand{\C}{\ensuremath{\text{\sf R}_{25}}}
\newcommand{\CC}{\ensuremath{\text{\sf R}_{50}}}
\newcommand{\CCC}{\ensuremath{\text{\sf R}_{75}}}
\newcommand{\CCCC}{\ensuremath{\text{\sf R}_{100}}}

\title{Parameter Effects in ReCom Ensembles}
\renewcommand{\thefootnote}{\fnsymbol{footnote}}

\author{
Kristopher Tapp\thanks{Professor of Mathematics, Saint Joseph's University}\\
\and
Todd Proebsting\thanks{Emeritus Professor of Computer Science, University of Arizona} \\
\and
Alec Ramsay\thanks{Dave's Redistricting}\\
}

\renewcommand{\thefootnote}{\arabic{footnote}}

\begin{document}
\maketitle

\begingroup
\renewcommand{\thefootnote}{}
\footnotetext[0]{Links to input data, code to generate, score, and analyze ensembles, our ensembles and scores and additional tables and plots, are noted in Supplementary Materials.
We thank 
Nick Stephanopoulos for his feedback as the first consumer of these ensembles,
Moon Duchin for her encouragement and guidance, and
Dave's Redistricting for sharing their data with us.}
\endgroup

\section{Introduction}
Ensemble analysis is a prominent technique for studying fair redistricting and for arguing to overturn biased maps in court.  The idea is to create an ensemble of thousands or millions of random maps in order to establish a baseline range for the behavior of maps drawn without partisan or racial intent. 

However, there is no such thing as ``\emph{the} ensemble'' for a given state and chamber.  Instead, there are multiple algorithms in common use, and each of them has several parameters whose settings could affect the metrics computed over those ensembles, possibly in unanticipated ways.  In advance of the next census cycle, we believe there is value in more systematically understanding the possible effects of these modeling choices, so that practitioners can be better informed when making arguments based on ensemble statistics.

In this paper, we focus on the ReCom algorithm, including its reversible variant.  We compare ReCom ensembles built with various parameter settings in multiple states (FL, IL, MI, NC, NY, OH, and WI) and legislative chambers (congressional, upper, and lower legislative).  More specifically, we study the extent to which its parameter settings can affect the statistics computed over an ensemble for commonly used scores related to compactness, partisan bias, competitiveness, and the opportunity for minority representation.  

\begin{figure}[bht!]\centering
\includegraphics[width=\textwidth]{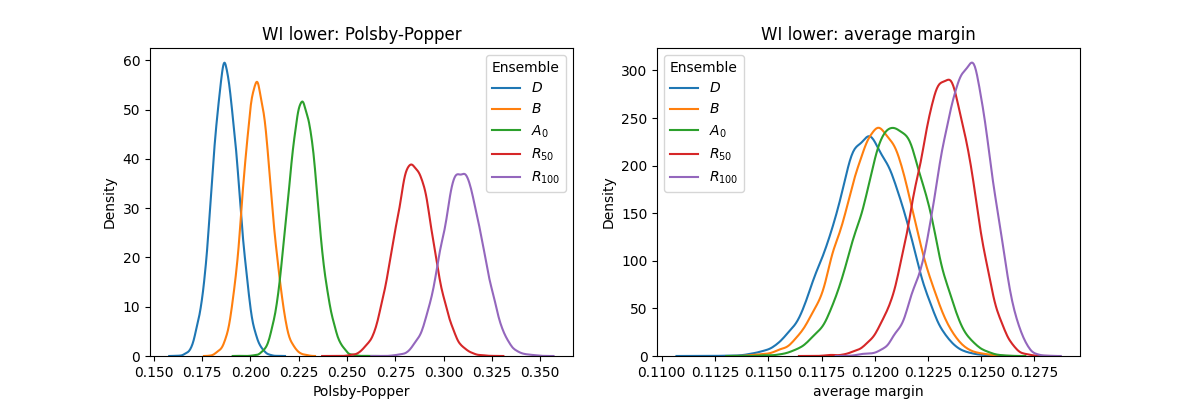}
\caption{kde plots for multiple ReCom variants (to be described later) with respect to a compactness score (left) and a competitiveness score (right) in the lower legislative chamber of Wisconsin.}\label{F:two_kde}
\end{figure}

As an example, while ReCom's parameter settings are known to affect compactness scores, Figure~\ref{F:two_kde} shows that they similarly affect a measure of competitiveness.  Below, we explain this phenomenon, which is surprisingly consistent over the states and chambers.   

But before comparing ensembles built using different parameter settings, one must first ensure the chain lengths are long enough that each individual ensemble yields repeatable statistics, independent of the initial seed plan.  A second goal of this paper is to introduce novel techniques for studying whether samples from these algorithms are sufficient to provide accurate statistics.
\section{Previous Work}
In the current census cycle, the algorithms used in court to generate ensembles have been based on spanning tree methods, including the ReCom algorithm introduced in~\cite{DeFord_Duchin_Solomon_2020}, its reversible variants (\cite{Autry_Carter_Herschlag_Hunter_Mattingly_2023} and~\cite{Cannon_Duchin_Randall_Rule_2022}), and the sequential Monte Carlo (SMC) algorithm introduced in~\cite{McCartan_Imai_2023}.  Here we briefly overview previous methods of comparing ensembles and determining whether a chain length is sufficient to produce a good ensemble.

The authors of~\cite{Fifield_Imai_Kawahara_Kenny_2020} and~\cite{McCartan_Imai_2023} compare ensembles for toy model states and for PA congressional maps, relying on the Gelman-Rubin $\hat{R}$-statistic to assess convergence.

The authors of~\cite{Cannon_Duchin_Randall_Rule_2022} introduced the RevReCom (reversible recombination) algorithm and compared its compactness statistics to other variants of ReCom.  They use the \emph{earth-mover} distance to analyze convergence for a 7-by-7 grid and also for PA and VA congressional plans.

There are dozens of reports and affidavits based on ensemble analysis, many of which use some sort of multi-start heuristic. In other words, they verify that the answers don't change much when the ensemble is redrawn from different random seeds.  In particular, we will borrow from ~\cite{Clelland_Colgate_DeFord_Malmskog_Sancier-Barbosa_2021}, which uses the Kolmogorov-Smirnov distance to assess convergence for Colorado congressional plans.

We will also assess convergence by measuring repetition effects, which borrows from~\cite{Cannon_DeFord_Duchin_2024}, whose authors found problematic levels of repetition in SMC ensembles; that is, a large portion of the maps in an SMC ensemble can have a large portion of their districts in common.  

\section{Our Contributions}
Most of the above-cited papers compare pairs of ensembles with respect to only one or two scores (usually the number of Democrat seats and/or a compactness measure) in an individual state.  To our knowledge, our paper is the first large-scale study to address these issues systematically across many states, chambers, ensemble types, and many scores.  This large amount of data allows us to offer the following contributions:
\begin{enumerate}
    \item We introduce novel methods of gauging whether the chain length is long enough to produce statistically good ensembles.  Specifically,
    \begin{enumerate}
        \item We employ results from MCMC theory to approximate effective samples sizes from autocorrelation measurements.
        \item We introduce new measurements of ensemble redundancy.
    \end{enumerate}
    \item We draw several conclusions about the extent to which the parameters affect the scores.  For example,
    \begin{enumerate}
        \item Changing the population tolerance has virtually no effect on any of the scores.
        \item Changing the other parameters can have a significant effect on \emph{partisan} and \emph{minority opportunity} statistics, but in a manner that is inconsistent across the states and chambers, and in some cases is a wash on average.
        \item Changing the parameters has a modest but significant effect on one \emph{competitiveness} score in a manner that is surprisingly consistent across the states and chambers.
    \end{enumerate}
\end{enumerate}

In the following sections, we discuss our methodology, how we know our ensembles converged, and, finally, the effects of parameter settings on the resulting scores.
\section{Methodology}

\subsection{Sample states and chambers}
We studied a cross-section of seven states: three balanced politically (NC, MI, and WI), two unbalanced towards Democrats (NY and IL), and two unbalanced towards Republicans (FL and OH).  For each of those, we generated congressional and state upper and lower house ensembles. 

\subsection{Input data and shapes}
We used precincts (Census VTDs) for our analysis and input data (GeoJSONs and adjacency graphs) from Dave’s Redistricting (DRA) noted in Supplementary Materials. We used the 2020 Census total population dataset (\verb|T_20_CENS|),  the 2020 VAP and CVAP datasets (\verb|V_20_VAP| and \verb|V_20_CVAP|, respectively), the 2016-2020 election composite dataset (\verb|E_16-20_COMP|),\footnote{Virtually all of the election data in DRA comes from partners, especially the \href{https://election.lab.ufl.edu/precinct-data/}{Voting and Election Science Team (VEST)} and \href{https://redistrictingdatahub.org/data/about-our-data/election-results-and-precinct-boundaries/}{Redistricting Data Hub}. Most of the data in the 2016-2020 timeframe came from VEST. See \href{https://davesredistricting.org/maps\#aboutdata}{DRA’s About Data page} for specific sources.} and DRA’s simplified shapes which derive from the Census Bureau's 2020 Tiger Line Shapefiles.  

\subsection{Subsampling}
For each chamber (congress, upper state legislative, lower state legislative) of each of seven states (FL, IL, MI, NC, NY, OH, WI), we create several ensembles that correspond to various parameter settings in the $\ReCom$ algorithm, including the reversible version.  More specifically, in each case, we use MGGG’s frcw.rs\footnote{We used a slightly modified, not-yet-published version.}
to build an ensemble of $\num{20000}$ plans that are formed by subsampling as follows:
\begin{itemize}
    \item We subsample every $\num{2500}^\text{th}$ plan in $50$ million step chain, for everything except reversible ReCom.
    \item For reversible ReCom, we subsample every $\num{50000}^\text{th}$ plan in a $1$ billion step chain.
\end{itemize}

Although some authors including~\cite{Geyer_1992} argue in favor of continuous observations, which means saving \emph{every} plan in the chain, subsampling is advantageous for creating a final product of manageable size.  While a subsample could never be better than the full chain, we will show that our approach yields ensembles of manageable size that are good enough to yield very accurate repeatable statistics.  In any case, it is self-evident that our approach builds a far better ensemble than would the approach of continuously observing a chain of length $\num{20000}$ to build an ensemble of size $\num{20000}$.

Subsampling is particularly useful for \emph{reversible} ReCom chains because of their high level of repetition (which is why we use longer chains for reversible ReCom).  The average number of steps for which a plan is repeated (before changing to a genuinely different plan) depends on the state and chamber but is in the thousands (see Table 1 of~\cite{Cannon_Duchin_Randall_Rule_2022}).  Therefore, subsampling offsets the repetition while preserving the distribution (whereas recording only the unique plans would change the distribution, undermining the whole purpose of the reversible ReCom algorithm).

\subsection{\ReCom Variants}\label{S:ensembles}
For each state and chamber combination, we generate the following ensembles.  Unless otherwise noted, the population tolerance is set to $.01$ for congressional and $.05$ for legislative chambers.
\begin{itemize}
    \item $\RA_0,...,\RA_4$: Five ensembles using different initial seed maps, all with the default algorithm that is labeled as \ReComA in~\cite{Cannon_Duchin_Randall_Rule_2022}.  More precisely, one step of this algorithm alters a map as follows:
    \begin{itemize}
        \item Select an adjacent pair of districts by choosing a random cut edge.
        \item Use a random minimum spanning tree (RMST) algorithm to draw a random  spanning tree of the merged district pair.
        \item If there are any cut edges that are balanced to within the population tolerance, randomly choose one and remove it to re-split the region into two districts.  Otherwise, return to the previous step (draw a new tree).
    \end{itemize}
    \item $\RB$, $\RC$, $\RD$: One ensemble for each of the variants labeled as \ReComB, \ReComC, and \ReComD in~\cite{Cannon_Duchin_Randall_Rule_2022}.  Compared to \ReComA,
    \begin{itemize}
        \item \ReComB selects which district pair to merge uniformly among all adjacent pairs (rather than via the choice of a random cut edge).
        \item \ReComC draws spanning trees with a uniform (random) spanning tree (UST) algorithm rather than an RMST algorithm.
        \item \ReComD makes both of these modifications.
    \end{itemize}
    \item \RevReCom: One ensemble from the reversible ReCom algorithm that was designed to target the spanning tree distribution; see~\cite{Cannon_Duchin_Randall_Rule_2022} for details.    
    \item $\C$, $\CC$, $\CCC$, $\CCCC$ : Ensembles in which the spanning trees are drawn with the county-aware version of an RMST algorithm with region-awareness strength (surcharge) settings of $0.25$, $0.50$, $0.75$ and $1.00$.  For example, a surcharge of $0.75$ means that each spanning tree is constructed by asking the RMST algorithm to find a minimal spanning tree with respect to edge weights uniformly chosen from $[0,1]$ for \emph{intra}-county edges and from $[.75,1.75]$ for \emph{inter}-county edges.  The higher the surcharge, the more the plans in the ensemble respect counties.
    \item $\popm$, $\popp$: Ensembles in which the population tolerance is decreased (to $.005$ for congressional and $.025$ for legislative) for $\popm$ and increased (to $.015$ for congressional and $.075$ for legislative) for $\popp$.
\end{itemize}
In total, we have $15$ ensembles for each of $7\cdot 3$ state and chamber combinations, for a total of $315$ ensembles.

There are a few previously observed differences between these algorithms with respect to compactness scores.  For example, \ReComA produces more compact maps than \ReComB, presumably because a long boundary between a pair of districts is more likely under \ReComA to be selected for merging, and hence replaced with a different and likely shorter boundary.  Furthermore, RMST algorithms tends to produce more compact plans than UST algorithms; see~\cite[Section 3]{Tapp_2025} for a partial explanation.

\subsection{Scores}\label{S:Scores}
We compute several commonly used scores for each map in each ensemble.
\begin{itemize}
    \item Three \emph{compactness} scores: \textbf{Reock}, \textbf{Polsby-Popper}, and \textbf{cut edges}; see for example~\cite{Duchin_Tenner_2023} for definitions.  The Reock and Polsby-Popper scores (which measure the compactness of single districts) are averaged over the districts.
    \item Four \emph{partisan bias} scores: \textbf{Dem seats} (the number of seats likely to be won by Democrats), \textbf{efficiency gap} (the wasted votes differential between the Democrats and Republicans, as originally defined in~\cite{Stephanopoulos_McGhee_2015}), \textbf{geometric seat bias ($\beta$)} (the failure of the seats-votes curve to be symmetric about $(0.5,0.5)$ defined in~\cite{Katz_King_Rosenblatt_2020}),\footnote{We use ``partisan bias'' to refer to this general category of metrics and ``geometric seat bias'' or simply ``seat bias'' to refer to this specific measure of it.} and the \textbf{mean-median} (the Democrat's mean vote share minus its median vote share across the districts). All four scores are defined with respect to the 2-party district-by-district vote shares (ignoring third party voters and nonvoters) using a blended election index (based on the two most recent presidential and two most recent senate races and the most recent Governor and Attorney General races). By convention, positive values for efficiency gap, geometric seats bias, and mean-median indicate Republican advantage and
negative values indicate Democratic advantage.
    \item Two \emph{competitiveness} scores involving the district-by-district vote margins, where for example, a $43\%$-to-$57\%$ district is considered to have a margin of $7\%$: the \textbf{average margin} (averaged over the districts), and the number of \textbf{competitive districts} (districts with less than $5\%$ margin). 
    \item Three \emph{minority opportunity} scores: \textbf{MMD black}, \textbf{MMD hispanic} and \textbf{MMD coalition}, which count the number Black-alone, Hispanic-alone, and ``coalition'' (Black and Hispanic-together) majority-minority districts.
    \item The number of \textbf{county splits}, defined as the number of counties that intersect multiple districts, counted with multiplicity so that, for example, a county that intersects $4$ districts contributes $3$ to the tally.  
\end{itemize}
In addition to those listed above, the Supplementary Materials include several dozen scores.

Table~\ref{tab:correlations} shows the correlation coefficients between pairs of scores.  In the following sections, we will discuss several interesting implications of this table. For now, notice that ``Dem seats'' and ``efficiency gap'' correlate strongly enough $(99\%)$ that we can consider them redundant and focus only on ``Dem seats'' in our analysis below.

\subsection{The distance between two ensembles}\label{S:metrics}
To ascertain whether an ensemble converged and to assess whether parameter variations produced statistically different results, we used several tests.

Given a pair $X,Y$ of random variables, there are several useful measurements of how close they are.  We are interested in the situation where $X$ and $Y$ represent the distributions of a score from Section~\ref{S:Scores} for two different ensembles of the same state and chamber.  These two ensembles might have different parameter settings (to study parameter effects) or might just have different initial seed plans (to study convergence).  

In any case, let $f_X,f_Y$ denote their probability mass functions, let $F_X,F_Y$ denote their cumulative distribution functions, let $\mu_X,\mu_Y$ denote their means, and let $\sigma_X$, $\sigma_Y$ denote their standard deviations.

\begin{itemize}
    \item The \textbf{mean-difference}: $d_\mu(X,Y) =|\mu_X-\mu_Y|$.
    \item The \textbf{earth-movers distance}: $\dem = \ell_1(F_X,F_Y)$.  Its name comes from the observation in~\cite{Rubner_Tomasi_Guibas_2000} that in the discrete case, $\dem(X,Y)$ can alternatively be defined as the total cost of the sequence of operations needed to convert $f_X$ into $f_Y$, where each operation has this form: subtract a value $v$ from $f_X(a)$ and add it to $f_X(b)$ at a cost of $v\cdot|a-b|$. This is also called the \emph{Wasserstein distance}.
    \item The \textbf{Kolmogorov-Smirnov distance}: $\dks(X,Y) = \ell_\infty(F_X,F_Y)$, where $\ell_\infty$ denotes the supremum metric.  Notice that $\dks(X,Y)\in[0,1]$, with the value $0$ indicating that $X$ and $Y$ are identically distributed, and the value $1$ indicating that $X$ and $Y$ have disjoint supports.  We find that the threshold $\dks>0.1$ roughly corresponds to the eyeball test (the difference is visually apparent when the kde plots are superimposed).
\end{itemize}

All of these metrics have been used by various authors in the redistricting setting.  For example, $\dem$ is used in~\cite{Cannon_Duchin_Randall_Rule_2022} to compare compactness scores of pairs of ensembles.  The authors of~\cite{Clelland_Colgate_DeFord_Malmskog_Sancier-Barbosa_2021} choose their chain length long enough to ensure empirically that 
\begin{equation}\label{E:Colorado}\mathbb{E}(\dks(X,Y)<.01),
\end{equation}
where $\mathbb{E}$ denotes the expected value, and $X,Y$ denotes the distributions of a score of interest for a pair of chains with different initial seed plans.  They observe that for i.i.d samples, Equation~\ref{E:Colorado} would require a sample size of $\num{15094}$, but since chains are subject to autocorrelation, a larger sample is needed; we will quantify this comment below.

The distances $d_\mu$ and $\dks$ have associated $p$-values representing the probability that the observed distance could have occurred by chance if $X$ and $Y$ had the same means (respectively, the same distributions), given their sample sizes and variances.  This is one advantage of $\dks$ over $\dem$.  Another advantage is that $\dks\in[0,1]$ is invariant under scaling ($X,Y\mapsto cX,cY$).

The \textbf{Gelman-Rubin $\hat{R}$-statistic}, which is the primary convergence test used in~\cite{McCartan_Imai_2023}, is essentially just a normalization of $d_\mu(X,Y)$; more precisely, when $X$ and $Y$ are samples from long chains,
$$\hat{R} \approx 1+\frac{d_\mu(X,Y)^2}{\sigma_X^2 + \sigma_Y^2}$$
In this situation, the threshold $\hat{R}<1.05$ that is used in~\cite{McCartan_Imai_2023, Fifield_Imai_Kawahara_Kenny_2020} becomes:
\begin{equation}\label{E:ALRAM}d_\mu(X,Y)\leq (.22)\sqrt{\sigma_X^2 + \sigma_Y^2},
\end{equation}
which is far weaker than the requirement $\dks<.01$ mentioned above.

When we compare the scores of an ensemble to those of the ``base'' ensemble, $\RA_0$, we will sometimes endow the measurements $d_\mu$ and $d_{KS}$ with a sign $\pm 1$ in the natural way, so that the measurement is positive when the ensemble has overall larger scores than $\RA_0$.  

\subsection{The distance between two ordered seat plots}\label{S:OSP}
\emph{Ordered seat plots} have become common in court cases.  For an ensemble of plans with $k$ districts, such a plot attempts to simultaneously illustrate the distributions of $k$ random variables, $X_1,...,X_k$ from left to right, where $X_i\in[0,1]$ is the distribution of the Democrat vote share of the $i^\text{th}$ reddest (most Republican) district of each plan in the ensemble.

It is useful to measure how sensitive such a plot is to the choice of ensemble.  If $X=\{X_i\}$ and $Y=\{Y_i\}$ come from two ensembles of the same state and chamber, we define the distance between the corresponding ordered seat plots as follows.
$$\dosp(X,Y) = \text{max}_{1\leq i\leq k}\left\{ \dks(X_i,Y_i) \
   \mid \mu_{X_i} \in[.45,.55]\text{ or }  \mu_{Y_i} \in[.45,.55]\right\}.$$
Rather than maximizing over \emph{all} seats, we're maximizing over the seats that are competitive (for at least one of the ensembles), which is the part of the plot that receives attention in court cases.  

\begin{figure}[bht!]
\centering
\scalebox{0.8}{
\begin{tikzpicture}
        \node (img) {\includegraphics[width=5in]{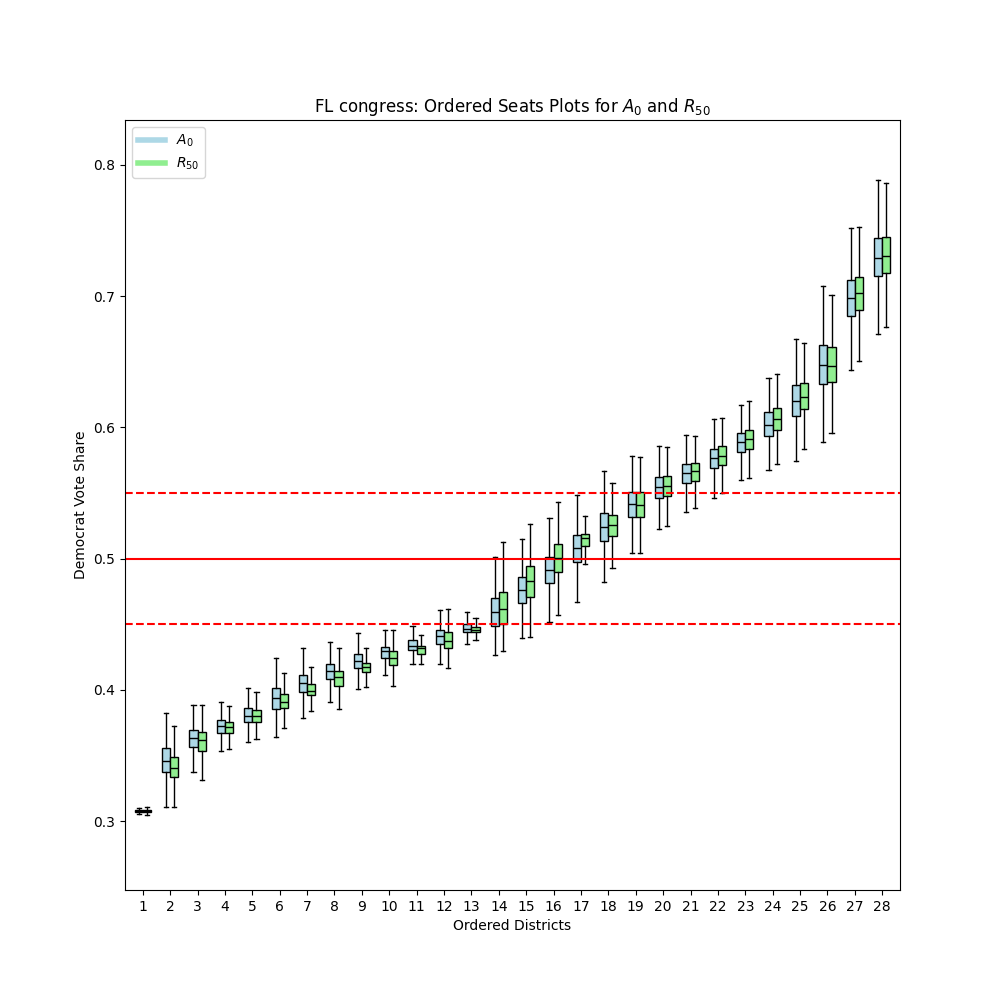}};
        \filldraw[fill=none, draw=cyan]
        ([xshift=2.48in, yshift=1.75in] img.south west) rectangle ++(.82in,1.03in); 
\end{tikzpicture}}
\caption{Ordered seats plots for two ensembles of FL congressional plans.}\label{F:OSP}
\end{figure}

For example, Figure~\ref{F:OSP} overlays the ordered seat plots for Florida congressional maps coming from two ensembles: $X = \RA_0$ and $Y = \CC$. Exactly $6$ of the $k = 28$ seats are competitive, namely seats $14-19$, as indicated by the cyan box.  Among these 6 competitive seats, the maximum $\dks$-distance is $.3$, which occurs at seat $17$.

\section{Chain length and convergence}
Although it is impossible to guarantee that our choices for the chain length and subsampling rate were sufficient to ensure adequate mixing, we employed several convergence heuristics to provide robust evidence that our samples were independent of the initial seed plans and sufficient for drawing statistical conclusions. 

The take-away is that, with the exception of the \RevReCom ensembles from certain states and chambers, all of our ensembles appear more than adequate for drawing very strong statistical conclusions. 
\subsection{Multi-start Evidence of Convergence}

Table~\ref{tab:multistart} provides strong evidence that our five \ReComA ensembles are independent of the initial seeds.  For each state and chamber combination, all pairs of the five ensembles (which differed in their initial seed plans) had $\dks<.026$ with respect to each score.  A similar table (not included) shows that, for each state, chamber and score, the Gelman-Rubin $\hat{R}$-statistic is less than $1.00045$ for all pairs of the five ensembles.

\subsection{Autocorrelation Evidence of Convergence}
It was only for the \ReComA algorithm that we created five ensembles using different seed plans.  Rather than doing so for the other parameter settings, in this section we use autocorrelation measurements to help indicate whether the sample is sufficient, without requiring multiple starts. 

Given an ordered list of scores $(s_1,...,s_n)$ from the steps of a Markov chain, the \textbf{lag-}$k$ \textbf{autocorrelation}, denoted $\gamma_k\in[-1,1]$, is defined as the correlation coefficient of the data points $\{(s_i,s_{i+k})\mid 1\leq i\leq n-k\}$.  When $\gamma_1$ is close to zero, this indicates that the scores are more like an i.i.d. sample of the distribution of scores of the full state space.  

Rather than viewing our ensemble as a subsample from a long Markov chain, we can view it as the output of a short Markov chain whose transition function involves taking $q$ steps of the long chain, where $q=\num{2500}$ for everything except \RevReCom chains where $q=\num{50000}$.  From this perspective, its lag-1 autocorrelation is comparable to the lag-$q$ autocorrelation of the long chain (more precisely, the former is an approximation of the latter based on a sample).  Thus $\gamma_1$ measures how close to mixing $q$ steps from the long chain come.

How close to zero must $\gamma_1$ be to provide evidence that $\num{20000}$ steps of the short chain are enough?  This can be quantified by relating the autocorrelation to the \textbf{Effective Sample Size}, denoted $\ESS$.  For example, if $\ESS=\num{10000}$, this intuitively means that our sample contains about as much information as an i.i.d. sample of size $\num{10000}$; this value is less than the actual size of $n=\num{20000}$ because the autocorrelation reduces our sample's effectiveness.  More precisely, $\ESS$ is defined as the value such that the \emph{standard error}, SE (meaning the standard deviation of the experiment of averaging a sample of size $n=\num{20000}$) is 
\begin{equation}\label{E:standarderror}\text{SE} = \frac{\sigma}{\sqrt{\ESS}},\end{equation}
where $\sigma$ is the standard deviation of the score's distribution on the full state space. 

There are several methods of approximating $\ESS$ that are based on the following formula (see, for example~\cite{Gelman_Carlin_Stern_Dunson_Vehtari_Rubin_2013} or~\cite{Geyer_book}):
\begin{equation}\label{E:Geyer}\ESS = \frac{n}{1+2\sum_{i=1}^\infty\gamma_i}.\end{equation}
Equation~\ref{E:Geyer} is only true under the idealized assumption that the chain is stationary; in other words, $\gamma_i$ here doesn't denote the correlation of two lists but rather of two random variables: the score of a draw from the stationary distribution, and the score $i$ steps later.  Unfortunately, if one substitutes our measured values of $\gamma_i$ into Equation~\ref{E:Geyer}, the sum is dominated by the noise of its tail.  Several strategies for addressing this issue are given in~\cite{Geyer_1992}.  We used the simple AR(1) model, which assumes that the series is geometric; that is, $\gamma_i = \gamma_1^i$ for all $i$.  In this case, Equation~\ref{E:Geyer} reduces to the following approximation:
\begin{equation}\label{E:Geyer2}\ESS \approx n\cdot\frac{1-\gamma_1}{1+\gamma_1}.\end{equation}

Table~\ref{tab:autocor} shows the value of $\gamma_1$ (in blue) and the corresponding approximation of $\ESS$ from Equation~\ref{E:Geyer2} (in red) for each ensemble with respect to several scores.

Are the $\ESS$ values in the table large enough to draw useful statistical conclusions?  To draw conclusions about the \emph{average} value of a score, it is straightforward to use Equation~\ref{E:standarderror} to determine the effective sample size needed to achieve a desired level of accuracy.  For example, achieving the accuracy of Equation~\ref{E:ALRAM} requires $\ESS\geq 21$.

Drawing conclusions about the whole distribution shape requires a larger sample. For this, a decent benchmark is the previously mentioned fact that an i.i.d. sample of size $\num{15094}$ is necessary to satisfy Equation~\ref{E:Colorado}.   With the exception of the columns for \RevReCom and the county-preserving ensembles, all of the $\ESS$ values throughout Table~\ref{tab:autocor} are larger than $\num{15094}$.  This would also have been (nearly) true of the county-preserving columns if we had omitted the ``county splits'' score from the list when we created the table; that is, the county splits score was the only score with respect to which the county preserving ensembles had poorer autocorrelations.

To our knowledge, the approximation of $\ESS$ given in Equation~\ref{E:Geyer2} has not been previously applied to the redistricting context, so it is important to study its accuracy.  One indication of its accuracy is the consistency between Tables~\ref{tab:multistart} and~\ref{tab:autocor}.  More precisely, Table~\ref{tab:autocor} shows that $\ESS>\num{15094}$ for the $\RA_0$ ensemble in each state and chamber, which is quite consistent with the fact that the values in Table~\ref{tab:multistart} are all around $.01$.  The ensemble redundancy analysis in the next section provides further supporting evidence of the ballpark accuracy of the approximation. 
\subsection{Ensemble redundancy}
Yet another way to gauge whether the chain length is long enough is to directly measure how different the plans are from each other.  In this section, we measure and report the amount of district-level redundancy in our ensembles.

An individual ensemble contains $S=\num{20000}$ plans, and each plan contains $k$ districts, where $k$ depends on the state and chamber.  Let $\mathcal{D}$ denote the list of all of the districts of all of the plans, so $|\mathcal{D}|=k\cdot S$.  If all of the plans of the ensemble were printed onto paper, and the $k$ districts of each were cut out with scissors, then $\mathcal{D}$ can be thought of as the resulting pile of paper districts.  

These districts are not all unique, so we define $n:\mathcal{D}\rightarrow\mathbb{N}$ so that $n(D)$ equals the number of elements of $\mathcal{D}$ that are identical to $D$ (which means they comprise the same set of precincts).  Two natural measurements of redundancy are:
$$\varphi_{\text{avg}} = \text{avg}\,n(\mathcal{D}),\qquad
\varphi_{\text{max}} = \text{max}\,n(\mathcal{D}),$$
where $n(\mathcal{D})=\left(n(D_1), \cdots ,n(D_{kS})\right),$ and $(D_1,...,D_{kS})$ is an enumeration of $\mathcal{D}$.  

The interpretation is that a random district of a random plan is repeated on average $\varphi_{\text{avg}}$ times, whereas $\varphi_{\text{max}}$ is the largest number of times that any district of any plan is repeated.

Table~\ref{tab:redundancy} reports the value of $\varphi_{\text{max}}$ and $\varphi_{\text{avg}}$ for each ensemble.  For example, in NC lower ($k=120$), the $\RevReCom$ ensemble has $\varphi_{\text{max}} = \num{766}$ and $\varphi_{\text{avg}} = 14.32$.  The later value means that a random district of a random plan is repeated on average $14.32$ times.  Although nonconsecutive repetition happens, we can roughly interpret this to mean that an average district takes about $14.32$ steps of the chain to disappear, which indicates that \RevReCom doesn't tend to refresh all of the districts between subsequent samples, in spite of the fact that it samples only every $\num{50000}^{th}$ plan.     

Contrast this with the $\CCCC$ ensemble of NC lower, which has a much lower $\varphi_{\text{avg}} = 1.86$ and a much higher $\varphi_{\text{max}} = \num{18651}$.  The latter value means that there is a single district that appears in most of the plans.  In this case, and for other high values of $\varphi_{\text{max}}$ in the county-preserving columns of the table, the repeated district is comprised of one or more adjacent counties that together fall within the desired population threshold\footnote{In the case of highly-repeated districts in county-preserving ensembles, and also in the case of mildly-repeated districts in other ensembles, the repetition is not necessarily in successive samples.  More than might be expected, districts disappear and reappear later in the chain.}.  Thus, the repetition is not necessarily a bug, but rather matches the way that a human map maker would probably insist on keeping whole any district-sized county (or adjacent counties).
 
These redundancy measurements complement the autocorrelation measurements of the previous section, with each approach capable of detecting things that the other misses.  For example, $\varphi_{\text{max}}$ would detect that a single district was repeated in all maps of the ensemble, but if the other districts were well-mixed, then the autocorrelation scores wouldn't notice.  On the other hand, $\varphi_{\text{max}}$ would be low for an ensemble built from a chain that made only minor alterations to district boundaries (like a flip-chain), whereas the autocorrelations of specific scores could detect the superficiality of this type of district diversity.

\section{Parameter Effects}
In this section, we employ the metrics from Section~\ref{S:metrics} to study the magnitude and significance of the effect of the ensemble types from Section~\ref{S:ensembles} on the scores from Section~\ref{S:Scores}.

More specifically, we will discuss the effect of changing the population tolerance (\popm, \popp), the ReCom variant (\RB, \RC, \RD), and the county surcharge (\C, \CC, \CCC, \CCCC).  In particular, we will discuss the effects of these changes on several types of scores: compactness, county-splitting, partisan bias, competitiveness, and minority opportunity.

To this end, Table~\ref{tab:OSP} reports the $d_\text{OSP}$ measurement defined in Section~\ref{S:OSP}, while
Tables~\ref{tab:dem_seats}-\ref{tab:MMD_coalition} show how the ensemble type affects the following scores: ``Dem seats,'' ``competitive seats,'' ``average margin,'' ``MMD Black, ``MMD Hispanic,'' and ``MMD coalition'' in all states and chambers.  The analogous tables for all other scores are available in Supplemental Material. 

The summary is that population tolerance does not have a material effect on any metrics.  As expected, the region surcharge and the ReCom variant affect compactness and the degree of county splitting.   
The region surcharge (and to a lesser degree the ReCom variant) frequently produced statistically and materially significant results for partisan bias, competitiveness, and the opportunity for minority representation.  Most of these effects are inconsistent across the states and chambers, but the effect on competitiveness is surprisingly consistent.
\subsection{Population tolerance does NOT affect the scores}
The tables show that changing the population tolerance does not have any significant or noticeable effect on any of the scores\footnote{An insignificant and very slight effect is that, in 18 of the 21 state and chamber combinations, the average cut edge counts of the ensembles is ordered as $\popp<\RA_0<\popm$; that is, decreasing the tolerance (tightening the population requirement) results in slightly less compact maps (more cut edges) for most states and chambers.  However, the differences are minuscule, and the superimposed histograms are nearly visually indistinguishable.}.  

\subsection{The ensemble type dramatically effects compactness and county splitting scores}
\begin{figure}[bht!]\centering
\includegraphics[width=4in]{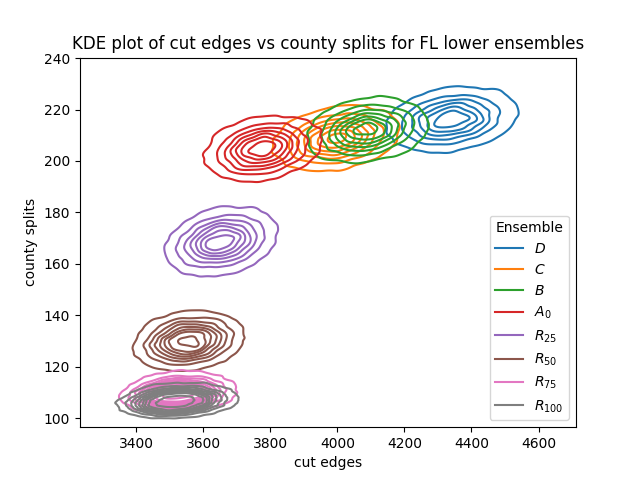}
\caption{County splits vs cut edges for FL lower (with $k=120$ districts).}\label{F:FL_kde}
\end{figure}

Unsurprisingly, both the ReCom variant and the county surcharge have dramatic effects on the compactness and county splitting scores, as illustrated by 
Figure~\ref{F:FL_kde}.  The figure is specific to the lower legislative chamber of Florida, but the corresponding plots for all of the other states and chambers are very similar.  The only qualitative difference is that, in many states and chambers, the $\RB$ and $\RC$ blobs are less separated than in Figure~\ref{F:FL_kde} and, in fact, are often indistinguishable.

Table~\ref{tab:correlations} shows a correlation of $0.81$ between ``cut edges'' and ``county splits,'' but Figure~\ref{F:FL_kde} indicates that this single number fails to tell the story.  More specifically, the figure indicates that the causation goes in both directions.  First, as the algorithm is asked to more stringently preserve counties ($\RA_0\rightarrow \C\rightarrow\CC\rightarrow\CCC\rightarrow\CCCC$), the resulting plans become more compact (fewer cut edges), presumably because the districts partially inherit the nice compact shapes of the counties.  Second, as we switch to algorithms that tend to produce less compact plans ($\RA_0\rightarrow\{\RB, \RC\}\rightarrow\RD$), the resulting maps split more counties, perhaps because districts with more wiggles and tentacles are prone to crossing more county lines.  The figure illustrates the two distinct slopes corresponding to these two directions of causality, so a single regression line would be a poor model here.  

It is important to recognize that the score one chooses to use to evaluate compactness can sometimes make a difference. The ``Polsby-Popper'' score aligns fairly well with ``cut edges'', but the ``Reock'' score sometimes does not.  For example, in some states and chambers, high region surcharges ($\CCC$ and $\CCCC$) move the Polsby-Popper score by many standard deviations relative to $\RA_0$, while Reock scores move by far less. 

\subsection{The ensemble type significantly but inconsistently affects partisan scores}
The county surcharge significantly affects partisan bias scores in many states and chambers.  For example, Table~\ref{tab:OSP} shows that $\dosp(\RA_0,\CCCC)$ has a maximum value of $0.47$ (WI lower) and an average value of $0.23$ over the states and chambers.  These values are high enough to indicate that there are visually obvious differences between the ordered seat plots.  

Table~\ref{tab:dem_seats} indicates that the average number of Democrat seats for the \RA and $\CCCC$ ensembles can differ by more than half a seat.  The maximum such difference is $0.74$ for the lower chamber of NC (with $k=120$).  The direction of the change is inconsistent -- a higher county surcharge helps the Democrats in some states and chambers and hurts them in others.  Averaged over the states and chambers that we considered, it helps them a tiny bit. 

The effect of the ReCom variant on partisan bias scores is overall milder than the effect of the county surcharge, but it is nevertheless significant in some states and chambers.  For example, for the lower chamber of NY, the $\RD$ ensemble has about half a seat more Democrat seats than $\RA_0$.  The direction of the change is inconsistent over the states and chambers, and on average, it’s a wash. 

The other partisan bias scores tell fairly similar stories.   The ``Dem seats'' aligns well with ``efficiency gap,'' but in some states and chambers, it aligns poorly with the ``geometric seat bias'' and ``mean-median'' scores.  For example, in NY congress, the ``Dem seats'' score indicates that a higher county surcharge hurts the Democrats, while the ``geometric seat bias' score indicates the opposite.  This lack of alignment is perhaps unsurprising, since the scores are designed to measure different things.

\subsection{The ensemble type significantly affects minority opportunity}
Both the county surcharge and the ReCom variant have significant effects on the opportunity for minority representation, as illustrated by Tables~\ref{tab:MMD_black}-\ref{tab:MMD_coalition}.  For example, changing from $\RA_0$ to $\CCCC$ can increase the average number of black MMDs by as much as $0.4$ (IL lower and NY upper), while changing from $\RA_0$ to $\RD$ can decrease the number of black MMDs by as much as $0.44$ (NY lower).

The magnitude of the effect is a bit lower for the MMD Hispanic score and is a bit higher for the MMD coalition score, but all of the changes are by less than 1 district.  

When changing from $\RA_0$ to $\CCCC$, the direction of the resultant change to the number of MMDs is inconsistent across the minority groups and the states and chambers.  However, when changing from $\RA_0$ to $\RD$, the effect is very consistent: $\RA_0$ has more average MMDs than $\RD$ for all minority groups and almost all states and chambers.  This is perhaps unsurprising.  Compared to $\RD$, the ensemble $\RA_0$ has districts that are a bit more compact and hence have slightly cleaner overlaps with regions of minority concentration.  The ensemble $\CCCC$ is even more compact, and in some state chambers, it might be too compact, resulting in a smaller number of packed MMDs rather than a larger number of barely-$50\%$ MMDs.     

\subsection{The ensemble type significantly and consistently affects competitiveness scores}
Both the county surcharge and the ReCom variant can have significant effects on measures of competitiveness.

One measurement of competitiveness is a simple count of ``competitive seats.''  Table~\ref{tab:competitive} shows that $\RA_0$ and $\CCCC$ can differ by as many as almost 3 competitive seats (WI lower), whereas $\RA_0$ and $\RD$ can differ by almost 1 competitive seat (FL congress).  These differences are inconsistent across the states and chambers, so the averaged differences are slight.

A second measurement of competitiveness is the ``average margin.''  The effect of the ensemble type on this score is mild, but (perhaps surprisingly) fairly consistent across the states and chambers.  For example, the $\RA_0$ ensemble for WI congressional ($k=8$) has an average margin of $8.03\%$, which means that the winning party in a random district of a random map of the ensemble receives $58.03\%$ of the vote share on average.  Table~\ref{tab:avg_margin} shows that the $\CCCC$ ensemble increases this score by $.86$ to $8.89\%$, while the $\RD$ ensemble decreases it by $.25$ to $7.78\%$.  

\begin{figure}[bht!]\centering
\includegraphics[width=3in]{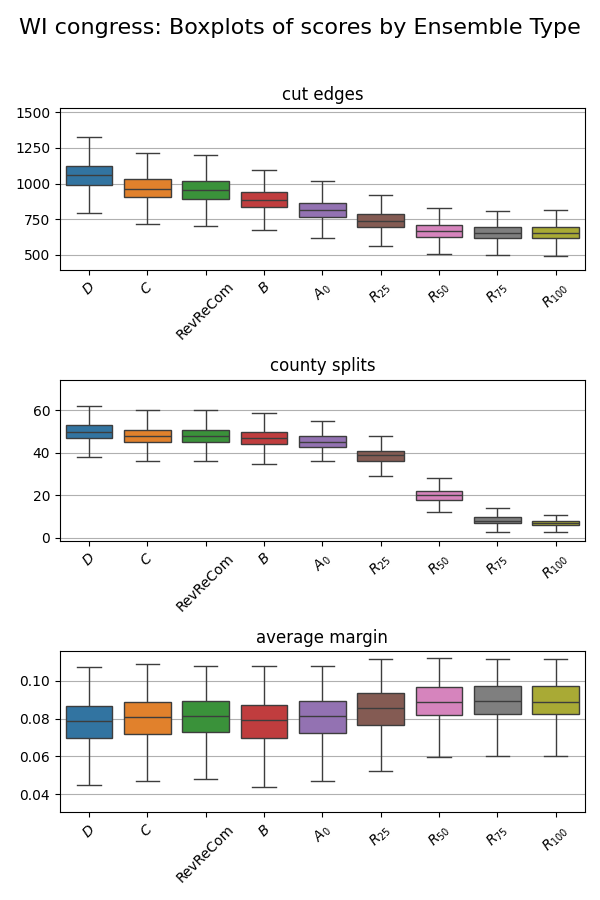}
\caption{Box and whisker plots for WI congress (with $k=8$ districts).}\label{F:box_whisker}
\end{figure}

These changes are modest, but their direction is fairly consistent.  Figure~\ref{F:box_whisker} shows for WI congress something that is roughly true for all states and chambers; namely, that  the ``average margin'' score is increasing with respect to the same ordering of the ensembles for which ``cut edges'' and ``county splits'' scores are decreasing:
$$\RD<\RC\approx \RevReCom\approx\RB<\RA_0<\C<\CC<\CCC\approx\CCCC.$$

This consistency is partially explained by correlations.  Table~\ref{tab:correlations} shows that the average margin score correlates with the compactness and county scores in a manner that is consistent across all of the states and chambers: maps that are more compact and/or have fewer county splits tend to have larger average margins.

What is special about the score AM = ``average margin'' that makes it correlate more consistently with compactness than do other scores like, say, ``Dem seats''?  To answer this, notice that (under the simplifying assumption that the districts all have equal turnout), AM equals the portion of all votes that are \emph{surplus votes}, which we can separate as $\text{AM} = S_D + S_R,$
where $S_D$ (respectively $S_R$) represent the portion of the total votes that are surplus votes and are in districts won by the Democrats (respectively the Republicans).  But the statewide Democrat vote share, $V_D$, can be expressed as:
$V_D = .5 + S_D - S_R$, so we get:
$$\text{AM} =  S_D + S_R = (V_D-.5)+2 S_R = (.5-V_D) + 2 S_D.$$

Suppose for specificity that Democrat voters are in the minority with $V_D=.45$.  In the limiting non-compact case, each district would spread thin tentacles throughout the state and inherit a vote share of about $.45$, giving all of the seats to the Republicans.  As the compactness increases, the ``Dem share'' score would behave in a complicated not-necessarily-monotonic manner that depends on the political geometry of the specific state.  For example, it might increase as the increasingly compact districts intersect with Democrat strongholds, but then decrease as too much intersection causes packing.  On the other hand, the formula $\text{AM} = .05 + 2 S_D$ shows that AM would increase monotonically.        

\subsection{Is there a \RevReCom proxy?}
How close are the \RevReCom ensembles to the other ensembles?  Which other ensembles are closest?  

These questions are important for two reasons.  First, \RevReCom is the only ensemble for which we can precisely describe the distribution from which it samples, namely the so-called \emph{spanning tree distribution}, so it is important to know how close the other variants come to sampling from the same distribution.  Second, generating a good \RevReCom ensemble is computationally expensive, so it would be useful if a computationally cheaper alternative could serve as a good proxy.     

\begin{figure}[bht!]\centering
\includegraphics[width=3.8in]{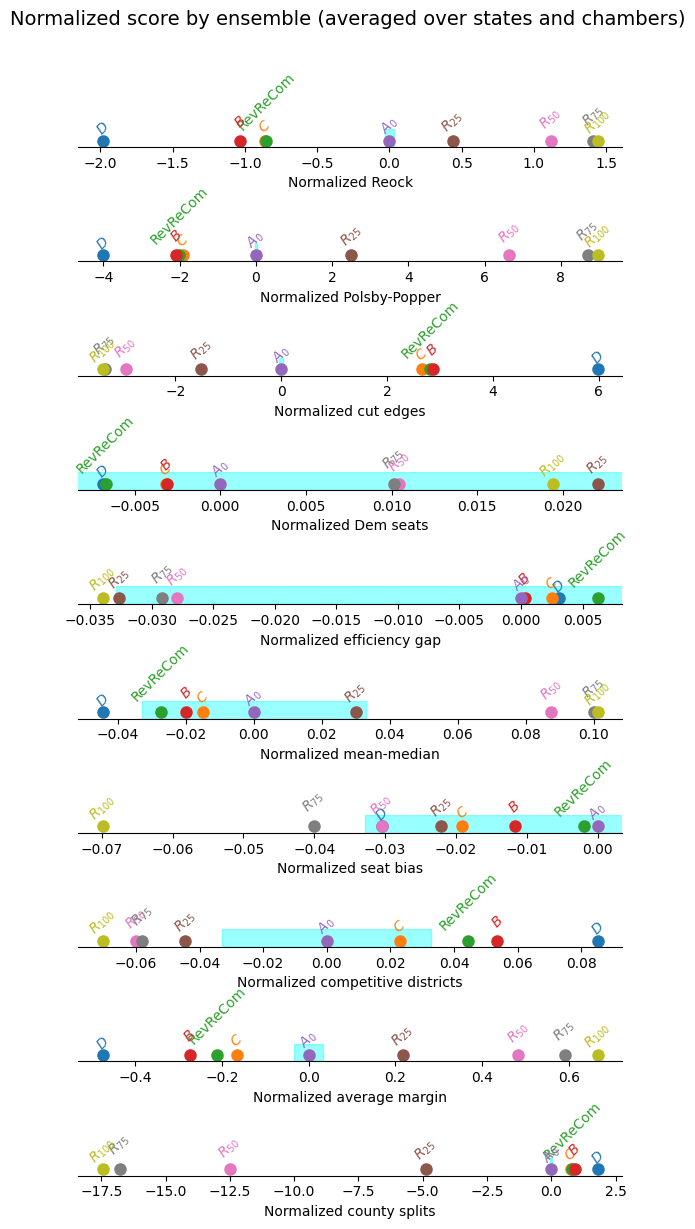}
\caption{\textbf{Score orders}: The position of an ensemble represents how many standard deviations is lies from $\RA_0$ with respect to that score.  Each cyan box shows the range $[-.033,.033]$; positions outside of the box are significantly different from $\RA_0$ at $p$-value $\leq 0.001$.
}\label{F:dot_plot}
\end{figure}

Figure~\ref{F:dot_plot} shows how close each ensemble is to \RevReCom with respect to each score.  The figure is averaged over the states and chambers, but the analogous plots for individual states and chambers show similar patterns and are available in the Supplementary Materials.  With respect to most scores, \RevReCom is close to and often between \RB and \RC, but the pattern isn't quite consistent enough to consider either as an adequate proxy across all of the scores, states, and chambers.  

\section{Conclusion}
This paper provides the first comprehensive, large-scale analysis of how the \ReCom algorithm parameters affect ensemble statistics across multiple states, chambers, and scoring metrics. Through a systematic study of 315 ensembles spanning seven states and three legislative chambers each, we have established several important findings for the redistricting community.

Our convergence analysis introduces novel techniques for assessing chain adequacy, including autocorrelation-based effective sample size estimates and ensemble redundancy measurements. These methods complement existing multi-start approaches and provide practitioners with robust tools for validating their ensemble quality. We demonstrate that, with appropriate chain lengths and subsampling, ReCom variants (except \RevReCom in certain states and chambers) produce statistically reliable ensembles that are independent of initial seed plans.

Regarding parameter sensitivity, we find that population tolerance changes have negligible effects on all metrics studied, suggesting practitioners can choose tolerance levels based on computational or legal constraints without affecting substantive conclusions. More importantly, we show that \ReCom variants and county-preservation settings do affect partisan bias, minority opportunity, and competitiveness statistics.  Some of these effects are surprisingly consistent across states and chambers. This finding has important implications for ensemble-based arguments about redistricting and suggests that modeling choices may introduce subtle biases.

These results provide essential guidance for practitioners using ensemble methods in litigation and policy analysis, helping ensure that modeling choices do not inadvertently undermine the credibility of redistricting arguments based on ensemble statistics.

\appendix
\section{Tables}

\begin{table}[ht]
\centering
\resizebox{\textwidth}{!}{
\begin{tabular}{llllllllllll}
\toprule
{} &   Reock & \makecell{Polsby \\ Popper} & \makecell{cut \\ edges} & \makecell{Dem \\ seats} & \makecell{efficiency \\ gap} & \makecell{mean \\ median} & \makecell{seat \\ bias} & \makecell{competitive \\ districts} & \makecell{average \\ margin} & \makecell{county \\ splits} & \makecell{MMD \\ coalition} \\
\midrule
Reock                 &    *1.0 &                       *0.73 &                  *-0.72 &                    0.01 &                        -0.02 &                      0.04 &                       -0.01 &                               -0.03 &                         0.26 &                      *-0.63 &                        0.06 \\
Polsby-Popper         &   *0.73 &                        *1.0 &                  *-0.91 &                    0.01 &                        -0.01 &                      0.05 &                       -0.01 &                               -0.04 &                        *0.33 &                      *-0.95 &                        0.06 \\
cut edges             &  *-0.72 &                      *-0.91 &                    *1.0 &                   -0.01 &                         0.01 &                     -0.05 &                         0.0 &                                0.05 &                       *-0.34 &                       *0.81 &                       -0.07 \\
Dem seats             &    0.01 &                        0.01 &                   -0.01 &                    *1.0 &                       *-0.99 &                     -0.24 &                      *-0.62 &                                 0.0 &                        -0.09 &                       -0.01 &                       -0.05 \\
\bottomrule
\end{tabular}
}
\caption{\textbf{Correlations}: the correlation coefficient between each pair of scores.  For each pair of scores, the correlation was computed separately for each state and chamber (using the scores from all maps of all ensembles), and then averaged over the states and chambers.  A $*$-symbol indicates that the sign was constant over all states and chambers.}
\label{tab:correlations}
\end{table}

\begin{table}[ht]
\centering
\resizebox{\textwidth}{!}{\begin{tabular}{lrrrrrrrrrr}
\toprule
{} &  Reock &  \makecell{Polsby \\ Popper} &  \makecell{cut \\ edges} &  \makecell{Dem \\ seats} &  \makecell{efficiency \\ gap} &  \makecell{mean \\ median} &  \makecell{seat \\ bias} &  \makecell{competitive \\ districts} &  \makecell{average \\ margin} &  \makecell{county \\ splits} \\
\midrule
FL 28  &   0.01 &                         0.01 &                     0.01 &                     0.01 &                          0.01 &                       0.01 &                         0.01 &                                 0.01 &                          0.01 &                         0.01 \\
FL 40  &   0.02 &                         0.01 &                     0.01 &                     0.01 &                          0.01 &                       0.02 &                         0.01 &                                 0.01 &                          0.02 &                         0.01 \\
FL 120 &   0.02 &                         0.01 &                     0.01 &                     0.01 &                          0.02 &                       0.01 &                         0.02 &                                 0.01 &                          0.01 &                         0.01 \\
IL 17  &   0.01 &                         0.01 &                     0.01 &                     0.01 &                          0.01 &                       0.01 &                         0.02 &                                 0.01 &                          0.01 &                         0.01 \\
IL 59  &   0.01 &                         0.01 &                     0.01 &                     0.01 &                          0.01 &                       0.01 &                         0.01 &                                 0.01 &                          0.01 &                         0.01 \\
\bottomrule
\end{tabular}
}
  \caption{\textbf{Multi-start differences}: the maximum $\dks$-distance between the 10 pairs of $\RA_0,...,\RA_4$, with all values rounded to 2 decimals.  The MMD scores are omitted to make the table fit on the page, but they also varied between $0.01$ and $0.02$.}
  \label{tab:multistart}
\end{table}

\begin{table}[ht]
\centering
\resizebox{\textwidth}{!}{\begin{tabular}{llllllllllll}
\toprule
{} &                                                      $\RA_0$ &                                                      $\popm$ &                                                      $\popp$ &                                                        $\RB$ &                                                        $\RC$ &                                                        $\RD$ &                                     \makecell{Rev \\ ReCom} &                                                         $\C$ &                                                        $\CC$ &                                                       $\CCC$ &                                                      $\CCCC$ \\
\midrule
FL 28  &  \makecell{\textcolor{blue}{0.06} \\ \textcolor{red}{17780}} &  \makecell{\textcolor{blue}{0.06} \\ \textcolor{red}{17864}} &  \makecell{\textcolor{blue}{0.06} \\ \textcolor{red}{17893}} &  \makecell{\textcolor{blue}{0.08} \\ \textcolor{red}{16990}} &  \makecell{\textcolor{blue}{0.05} \\ \textcolor{red}{18064}} &  \makecell{\textcolor{blue}{0.04} \\ \textcolor{red}{18418}} &   \makecell{\textcolor{blue}{0.98} \\ \textcolor{red}{227}} &  \makecell{\textcolor{blue}{0.07} \\ \textcolor{red}{17453}} &  \makecell{\textcolor{blue}{0.13} \\ \textcolor{red}{15316}} &  \makecell{\textcolor{blue}{0.14} \\ \textcolor{red}{15036}} &  \makecell{\textcolor{blue}{0.23} \\ \textcolor{red}{12642}} \\
\bottomrule
\end{tabular}
}
\caption{\textbf{Autocorrelation and effective sample sizes}.  BLUE: the maximum (over all scores except the MMD scores) of the lag-1 autocorrelation, rounded to 2 digits. RED: the corresponding estimated effective sample size.  The MMD scores are omitted here because some states and chambers do not have any MMD districts in their ensembles.}
\label{tab:autocor}
\end{table}

\begin{table}[ht]
\centering
\resizebox{\textwidth}{!}{\begin{tabular}{llllllllllll}
\toprule
{} &                                                    $\RA_0$ &                                                    $\popm$ &                                                    $\popp$ &                                                      $\RB$ &                                                      $\RC$ &                                                      $\RD$ &                                     \makecell{Rev \\ ReCom} &                                                        $\C$ &                                                        $\CC$ &                                                       $\CCC$ &                                                      $\CCCC$ \\
\midrule
FL 28  &  \makecell{\textcolor{blue}{372} \\ \textcolor{red}{1.03}} &  \makecell{\textcolor{blue}{614} \\ \textcolor{red}{1.03}} &  \makecell{\textcolor{blue}{239} \\ \textcolor{red}{1.03}} &  \makecell{\textcolor{blue}{243} \\ \textcolor{red}{1.02}} &  \makecell{\textcolor{blue}{322} \\ \textcolor{red}{1.03}} &  \makecell{\textcolor{blue}{204} \\ \textcolor{red}{1.02}} &   \makecell{\textcolor{blue}{325} \\ \textcolor{red}{2.76}} &   \makecell{\textcolor{blue}{529} \\ \textcolor{red}{1.04}} &   \makecell{\textcolor{blue}{3911} \\ \textcolor{red}{1.09}} &   \makecell{\textcolor{blue}{7167} \\ \textcolor{red}{1.13}} &   \makecell{\textcolor{blue}{7557} \\ \textcolor{red}{1.14}} \\
\bottomrule
\end{tabular}
}
\caption{\textbf{Ensemble Redundancy}.  BLUE: $\varphi_{\text{max}}$ = the maximum number of times that any district of any plan is repeated in the ensemble.  RED: $\varphi_{\text{avg}}$ = the average number of times that a random district of a random plan is repeated in the ensemble.}
\label{tab:redundancy}
\end{table}

\begin{table}[ht]
\centering
\resizebox{\textwidth}{!}{\begin{tabular}{lrrrrrrrrrr}
\toprule
{} &  $\RA_1$ &  $\popm$ &  $\popp$ &  $\RB$ &  $\RC$ &  $\RD$ &  $\C$ &  $\CC$ &  $\CCC$ &  $\CCCC$ \\
\midrule
FL 28   &     0.02 &     0.02 &     0.02 &   0.13 &   0.09 &   0.24 &  0.18 &   0.30 &    0.30 &     0.29 \\
FL 40   &     0.02 &     0.04 &     0.07 &   0.04 &   0.02 &   0.07 &  0.06 &   0.11 &    0.13 &     0.16 \\
FL 120  &     0.01 &     0.03 &     0.01 &   0.03 &   0.03 &   0.05 &  0.07 &   0.09 &    0.09 &     0.09 \\
IL 17   &     0.01 &     0.01 &     0.01 &   0.04 &   0.07 &   0.06 &  0.09 &   0.21 &    0.25 &     0.25 \\
IL 59   &     0.01 &     0.02 &     0.03 &   0.03 &   0.04 &   0.07 &  0.05 &   0.13 &    0.17 &     0.20 \\
IL 118  &     0.01 &     0.02 &     0.02 &   0.03 &   0.04 &   0.07 &  0.04 &   0.13 &    0.22 &     0.26 \\
MI 13   &     0.01 &     0.01 &     0.01 &   0.07 &   0.09 &   0.15 &  0.04 &   0.15 &    0.21 &     0.22 \\
\bottomrule
\end{tabular}
}
  \caption{\textbf{Ordered seat plot differences}: the maximum over the competitive seats of the $\dks$-distance (between that ensemble and the $\RA_0$ ensemble) for that seat's Democrat vote share.  The \RevReCom ensembles are omitted because some of their effective samples sizes are insufficient for accurate $\dks$ comparisons.}
  \label{tab:OSP}
\end{table}

\begin{table}[ht]
\centering
\resizebox{\textwidth}{!}{\begin{tabular}{lrlrllllllll}
\toprule
{} &  $\RA_1$ &        $\popm$ &  $\popp$ &           $\RB$ & \makecell{Rev \\ ReCom} &           $\RC$ &           $\RD$ &            $\C$ &           $\CC$ &          $\CCC$ &         $\CCCC$ \\
\midrule
FL 28   &     0.00 &           0.01 &     0.00 &  \textbf{-0.18} &          \textbf{-0.16} &  \textbf{-0.11} &  \textbf{-0.29} &   \textbf{0.42} &   \textbf{0.59} &   \textbf{0.56} &   \textbf{0.55} \\
FL 40   &     0.02 &          -0.01 &     0.02 &  \textbf{-0.09} &          \textbf{-0.07} &           -0.02 &  \textbf{-0.14} &   \textbf{0.07} &            0.04 &   \textbf{0.05} &    \textbf{0.1} \\
FL 120  &     0.01 &           0.03 &     0.03 &            0.01 &           \textbf{0.16} &            -0.0 &           -0.06 &  \textbf{-0.12} &  \textbf{-0.21} &  \textbf{-0.09} &   \textbf{0.07} \\
IL 17   &     0.00 &          -0.01 &    -0.00 &   \textbf{0.03} &           \textbf{0.03} &            0.01 &   \textbf{0.04} &  \textbf{-0.06} &  \textbf{-0.23} &  \textbf{-0.23} &  \textbf{-0.21} \\
IL 59   &    -0.01 &          -0.03 &     0.04 &           -0.01 &           \textbf{0.14} &    \textbf{0.1} &   \textbf{0.09} &           -0.03 &             0.0 &  \textbf{-0.07} &  \textbf{-0.15} \\
IL 118  &     0.01 &           0.03 &     0.00 &           -0.05 &                    0.01 &   \textbf{0.08} &            0.01 &  \textbf{-0.14} &  \textbf{-0.14} &  \textbf{-0.22} &  \textbf{-0.35} \\
MI 13   &    -0.00 &            0.0 &     0.00 &   \textbf{0.05} &           \textbf{0.07} &   \textbf{0.06} &   \textbf{0.09} &  \textbf{-0.04} &  \textbf{-0.16} &  \textbf{-0.26} &  \textbf{-0.27} \\
\bottomrule
\end{tabular}
}
  \caption{\textbf{Dem seats}: the mean-difference for the number of Democrat seats between the given ensemble and the $\RA_0$ ensemble.  Positive numbers indicate that the ensemble advantages Democrats by that fraction of a seat, while negative numbers advantage Republicans.  Bold indicates that the difference is significant at p-value $<.001$.}
  \label{tab:dem_seats}
\end{table}

\begin{table}[ht]
\centering
\resizebox{\textwidth}{!}{\begin{tabular}{lrllllllllll}
\toprule
{} &  $\RA_1$ &         $\popm$ &         $\popp$ &           $\RB$ & \makecell{Rev \\ ReCom} &           $\RC$ &           $\RD$ &            $\C$ &           $\CC$ &          $\CCC$ &         $\CCCC$ \\
\midrule
FL 28   &    -0.02 &           -0.04 &           -0.01 &   \textbf{0.46} &           \textbf{0.42} &    \textbf{0.3} &   \textbf{0.88} &           -0.03 &  \textbf{-0.09} &  \textbf{-0.11} &   \textbf{-0.1} \\
FL 40   &    -0.02 &           -0.02 &  \textbf{-0.06} &            0.03 &                    0.02 &            0.05 &   \textbf{0.16} &  \textbf{-0.14} &  \textbf{-0.31} &  \textbf{-0.37} &  \textbf{-0.43} \\
FL 120  &     0.05 &            0.09 &            -0.0 &   \textbf{0.11} &           \textbf{0.24} &   \textbf{0.14} &   \textbf{0.19} &   \textbf{0.45} &   \textbf{0.39} &   \textbf{0.28} &   \textbf{0.44} \\
IL 17   &    -0.00 &           -0.02 &           -0.01 &   \textbf{0.04} &           \textbf{0.04} &            0.02 &   \textbf{0.06} &  \textbf{-0.16} &            0.01 &   \textbf{0.06} &   \textbf{0.05} \\
IL 59   &    -0.04 &  \textbf{-0.12} &    \textbf{0.1} &   \textbf{0.08} &           \textbf{0.28} &   \textbf{0.07} &   \textbf{0.21} &  \textbf{-0.07} &   \textbf{0.13} &   \textbf{0.26} &   \textbf{0.37} \\
IL 118  &     0.00 &           -0.02 &            0.05 &   \textbf{0.25} &           \textbf{0.37} &    \textbf{0.1} &   \textbf{0.34} &   \textbf{0.23} &   \textbf{0.78} &   \textbf{1.43} &   \textbf{1.65} \\
MI 13   &     0.00 &           -0.01 &            -0.0 &  \textbf{-0.07} &          \textbf{-0.05} &  \textbf{-0.06} &  \textbf{-0.15} &            0.01 &   \textbf{0.07} &   \textbf{0.12} &    \textbf{0.1} \\
\bottomrule
\end{tabular}
}
  \caption{\textbf{Competitive seats}: the mean-difference for the number of competitive seats between the given ensemble and the $\RA_0$ ensemble.  Positive numbers indicate that the ensemble has more competitive seats than the $\RA_0$ ensemble.  Bold indicates that the difference is significant at p-value $<.001$.}
  \label{tab:competitive}
\end{table}

\begin{table}[ht]
\centering
\resizebox{\textwidth}{!}{\begin{tabular}{lrllllllllll}
\toprule
{} &  $\RA_1$ &         $\popm$ &        $\popp$ &           $\RB$ & \makecell{Rev \\ ReCom} &           $\RC$ &           $\RD$ &            $\C$ &          $\CC$ &          $\CCC$ &         $\CCCC$ \\
\midrule
FL 28   &     0.01 &            0.01 &           -0.0 &  \textbf{-0.15} &          \textbf{-0.13} &  \textbf{-0.11} &  \textbf{-0.29} &   \textbf{0.08} &  \textbf{0.17} &   \textbf{0.19} &   \textbf{0.19} \\
FL 40   &     0.00 &           -0.01 &  \textbf{0.02} &  \textbf{-0.05} &          \textbf{-0.01} &  \textbf{-0.04} &  \textbf{-0.11} &   \textbf{0.05} &  \textbf{0.09} &   \textbf{0.08} &   \textbf{0.08} \\
FL 120  &     0.00 &  \textbf{-0.01} &  \textbf{0.01} &  \textbf{-0.08} &          \textbf{-0.12} &  \textbf{-0.05} &  \textbf{-0.13} &   \textbf{0.03} &  \textbf{0.07} &    \textbf{0.1} &    \textbf{0.1} \\
IL 17   &     0.00 &             0.0 &           -0.0 &  \textbf{-0.02} &           \textbf{0.01} &   \textbf{0.01} &  \textbf{-0.02} &  \textbf{-0.01} &  \textbf{0.06} &   \textbf{0.11} &   \textbf{0.12} \\
IL 59   &     0.00 &            -0.0 &            0.0 &  \textbf{-0.07} &          \textbf{-0.06} &  \textbf{-0.04} &  \textbf{-0.12} &   \textbf{0.04} &  \textbf{0.09} &    \textbf{0.1} &    \textbf{0.1} \\
IL 118  &    -0.00 &             0.0 &            0.0 &  \textbf{-0.07} &          \textbf{-0.08} &  \textbf{-0.03} &   \textbf{-0.1} &   \textbf{0.01} &  \textbf{0.01} &  \textbf{-0.01} &  \textbf{-0.02} \\
MI 13   &     0.00 &             0.0 &           -0.0 &  \textbf{-0.03} &                   -0.01 &           -0.01 &  \textbf{-0.03} &            0.01 &  \textbf{0.02} &   \textbf{0.04} &   \textbf{0.05} \\
\bottomrule
\end{tabular}
}
  \caption{\textbf{Average margin}: the mean-difference for the average margin between the given ensemble and the $\RA_0$ ensemble with values converted to percents (multiplied by \num{100}).  Positive numbers indicate that the ensemble has a higher average margins than the $\RA_0$ ensemble.  Bold indicates that the difference is significant at p-value $<.001$.}
  \label{tab:avg_margin}
\end{table}

\begin{table}[ht]
\centering
\resizebox{\textwidth}{!}{\begin{tabular}{lrrrllllllll}
\toprule
{} &  $\RA_1$ &  $\popm$ &  $\popp$ &           $\RB$ & \makecell{Rev \\ ReCom} &           $\RC$ &           $\RD$ &            $\C$ &           $\CC$ &          $\CCC$ &         $\CCCC$ \\
\midrule
FL 120  &    -0.00 &    -0.01 &     0.01 &  \textbf{-0.17} &          \textbf{-0.36} &   \textbf{-0.1} &  \textbf{-0.28} &  \textbf{-0.05} &  \textbf{-0.13} &  \textbf{-0.17} &  \textbf{-0.17} \\
IL 59   &    -0.00 &    -0.02 &     0.01 &  \textbf{-0.12} &          \textbf{-0.13} &  \textbf{-0.04} &  \textbf{-0.17} &   \textbf{0.06} &   \textbf{0.12} &   \textbf{0.15} &   \textbf{0.15} \\
IL 118  &     0.01 &    -0.03 &     0.00 &  \textbf{-0.11} &            \textbf{0.1} &  \textbf{-0.05} &  \textbf{-0.15} &   \textbf{0.08} &   \textbf{0.21} &   \textbf{0.34} &    \textbf{0.4} \\
MI 38   &    -0.01 &    -0.00 &    -0.00 &  \textbf{-0.08} &          \textbf{-0.07} &  \textbf{-0.06} &  \textbf{-0.13} &  \textbf{-0.03} &  \textbf{-0.07} &  \textbf{-0.06} &  \textbf{-0.06} \\
MI 110  &     0.01 &    -0.00 &     0.00 &  \textbf{-0.06} &          \textbf{-0.14} &           -0.01 &  \textbf{-0.05} &  \textbf{-0.07} &  \textbf{-0.14} &  \textbf{-0.15} &  \textbf{-0.17} \\
NC 50   &     0.01 &     0.02 &     0.01 &  \textbf{-0.08} &                   -0.01 &  \textbf{-0.04} &  \textbf{-0.13} &   \textbf{0.06} &   \textbf{0.13} &   \textbf{0.15} &   \textbf{0.15} \\
NC 120  &     0.00 &     0.02 &     0.03 &   \textbf{-0.2} &           \textbf{0.14} &  \textbf{-0.12} &  \textbf{-0.36} &   \textbf{0.06} &   \textbf{0.13} &   \textbf{0.17} &   \textbf{0.17} \\
\bottomrule
\end{tabular}
}
  \caption{\textbf{MMD Black}: the mean-difference for the number of black MMDs between the given ensemble and the $\RA_0$ ensemble.  Bold indicates that the difference is significant at p-value $<.001$.  Only state and chamber combinations for which $\RA_0$ has more than $1$ MMD on average are included.}
  \label{tab:MMD_black}
\end{table}

\begin{table}[ht]
\centering
\resizebox{\textwidth}{!}{\begin{tabular}{lrlrllllllll}
\toprule
{} &  $\RA_1$ &         $\popm$ &  $\popp$ &           $\RB$ & \makecell{Rev \\ ReCom} &           $\RC$ &           $\RD$ &            $\C$ &           $\CC$ &          $\CCC$ &         $\CCCC$ \\
\midrule
FL 28   &    -0.00 &             0.0 &     0.00 &  \textbf{-0.07} &          \textbf{-0.03} &  \textbf{-0.02} &   \textbf{-0.1} &   \textbf{0.02} &   \textbf{0.04} &   \textbf{0.04} &   \textbf{0.04} \\
FL 40   &     0.01 &            0.01 &     0.02 &  \textbf{-0.11} &                   -0.01 &  \textbf{-0.03} &  \textbf{-0.16} &  \textbf{-0.13} &  \textbf{-0.29} &  \textbf{-0.33} &  \textbf{-0.36} \\
FL 120  &    -0.00 &  \textbf{-0.04} &    -0.01 &  \textbf{-0.07} &          \textbf{-0.17} &  \textbf{-0.04} &  \textbf{-0.09} &  \textbf{-0.09} &  \textbf{-0.11} &  \textbf{-0.03} &  \textbf{-0.05} \\
IL 59   &    -0.01 &           -0.01 &    -0.01 &  \textbf{-0.05} &          \textbf{-0.05} &  \textbf{-0.03} &  \textbf{-0.07} &           -0.01 &           -0.01 &             0.0 &            -0.0 \\
IL 118  &    -0.01 &            -0.0 &    -0.01 &  \textbf{-0.07} &          \textbf{-0.04} &           -0.03 &  \textbf{-0.12} &            0.02 &            0.01 &             0.0 &           -0.01 \\
NY 26   &     0.00 &            0.01 &     0.00 &  \textbf{-0.05} &          \textbf{-0.03} &  \textbf{-0.02} &  \textbf{-0.08} &   \textbf{0.13} &   \textbf{0.23} &   \textbf{0.23} &   \textbf{0.23} \\
NY 63   &    -0.01 &            -0.0 &     0.02 &  \textbf{-0.13} &          \textbf{-0.07} &  \textbf{-0.08} &  \textbf{-0.22} &   \textbf{0.03} &            -0.0 &   \textbf{0.05} &   \textbf{0.09} \\
\bottomrule
\end{tabular}
}
  \caption{\textbf{MMD Hispanic}: the mean-difference for the number of hisanic MMDs between the given ensemble and the $\RA_0$ ensemble.  Bold indicates that the difference is significant at p-value $<.001$.  Only state and chamber combinations for which $\RA_0$ has more than $1$ MMD on average are included.}
  \label{tab:MMD_hispanic}
\end{table}

\begin{table}[ht]
\centering
\resizebox{\textwidth}{!}{\begin{tabular}{lrrlllllllll}
\toprule
{} &  $\RA_1$ &  $\popm$ &        $\popp$ &           $\RB$ & \makecell{Rev \\ ReCom} &           $\RC$ &           $\RD$ &            $\C$ &           $\CC$ &          $\CCC$ &         $\CCCC$ \\
\midrule
FL 28   &     0.01 &     0.02 &           0.01 &  \textbf{-0.09} &           \textbf{0.03} &  \textbf{-0.07} &  \textbf{-0.19} &   \textbf{0.06} &   \textbf{0.17} &   \textbf{0.18} &   \textbf{0.18} \\
FL 40   &    -0.01 &    -0.02 &           0.03 &  \textbf{-0.08} &                    0.01 &  \textbf{-0.03} &  \textbf{-0.15} &   \textbf{0.07} &   \textbf{0.22} &   \textbf{0.22} &   \textbf{0.18} \\
FL 120  &     0.02 &     0.02 &           0.03 &  \textbf{-0.25} &          \textbf{-0.82} &  \textbf{-0.14} &   \textbf{-0.4} &            0.01 &           -0.02 &            0.03 &   \textbf{0.08} \\
IL 17   &     0.00 &    -0.00 &           0.01 &           -0.01 &           \textbf{0.04} &   \textbf{0.05} &            0.02 &   \textbf{0.05} &   \textbf{0.17} &   \textbf{0.19} &   \textbf{0.19} \\
IL 59   &    -0.01 &    -0.02 &          -0.01 &  \textbf{-0.05} &          \textbf{-0.07} &           -0.02 &  \textbf{-0.08} &   \textbf{0.08} &   \textbf{0.17} &   \textbf{0.19} &   \textbf{0.19} \\
IL 118  &    -0.00 &    -0.01 &          -0.01 &  \textbf{-0.11} &          \textbf{-0.35} &           -0.04 &  \textbf{-0.17} &   \textbf{0.28} &   \textbf{0.54} &   \textbf{0.64} &   \textbf{0.65} \\
MI 38   &    -0.01 &    -0.01 &           -0.0 &  \textbf{-0.08} &          \textbf{-0.06} &  \textbf{-0.06} &  \textbf{-0.13} &  \textbf{-0.06} &  \textbf{-0.15} &  \textbf{-0.16} &  \textbf{-0.16} \\
\bottomrule
\end{tabular}
}
  \caption{\textbf{MMD Coalition}: the mean-difference for the number of coalition MMDs between the given ensemble and the $\RA_0$ ensemble.  Bold indicates that the difference is significant at p-value $<.001$.  Only state and chamber combinations for which $\RA_0$ has more than $1$ MMD on average are included.}
  \label{tab:MMD_coalition}
\end{table}

\clearpage
\bibliographystyle{alpha}  
\bibliography{bibliography}  
\section*{Supplementary Materials}
The input data that we used is available at \url{https://github.com/dra2020/vtd_data}. The code to generate, score, and analyze ensembles is available at \url{https://github.com/proebsting/TBD}, \url{https://dra2020.github.io/rdapy/}, and \url{https://github.com/KrisTapp/Knobs}, respectively. The ensembles and scores themselves are available at \url{https://TBD-url}.
\end{document}